\documentclass[aps,floatfix,showpacs,twocolumn,superscriptaddress]{revtex4} 
\usepackage{amssymb,amsmath}
\usepackage{graphicx}
\usepackage{dcolumn}
\begin{document}

%\preprint{\parbox[t]{45mm}{\small ANL-PHY-11422-TH-2005, E2-2005-176, MPG-VT-UR 264/05}}

\title{Pair production and optical lasers}

\author{D.\,B.\ Blaschke}
\affiliation{Gesellschaft f\"ur Schwerionenforschung (GSI) mbH,
Planckstr.\ 1, D-64291 Darmstadt, Germany}
\affiliation{Bogoliubov Laboratory for Theoretical Physics, Joint Institute for
Nuclear Research, RU-141980, Dubna, Russia
}%

\author{A.\,V.\ Prozorkevich}
\affiliation{Saratov State University, RU-410026,
Saratov, Russia }

\author{C.\,D.\ Roberts} 
%\email[]{cdroberts@anl.gov} 
%\homepage[]{http://www.phy.anl.gov/theory/staff/cdr.html} 
\affiliation{Physics Division, Argonne National Laboratory, 
             Argonne, IL 60439-4843, U.S.A.} 
\affiliation{Institut f\"ur Physik, Universit\"at Rostock, D-18051 Rostock, 
Germany} 

\author{S.\,M.\ Schmidt}
\affiliation{Helmholtz-Gemeinschaft, Ahrstrasse 45, D-53175 Bonn, Germany}

\author{S.\,A.\ Smolyansky}
\affiliation{Saratov State University, RU-410026,
Saratov, Russia }

\date{\today}

\begin{abstract}
\rule{0ex}{3ex} 
Electron-positron pair creation in a standing wave is explored using a parameter-free quantum kinetic equation.  Field strengths and frequencies corresponding to modern optical lasers induce a material polarisation of the QED vacuum, which may be characterised as a plasma of $e^+ e^-$ quasiparticle pairs with a density of $\sim 10^{20}$cm$^{-3}$.  The plasma vanishes almost completely when the laser field is zero, leaving a very small residual pair density, $n_r$, which is the true manifestation of vacuum decay.  The average pair density per period is proportional to the laser intensity but independent of the frequency, $\nu$.
%, for $\nu \ll m$.  
The density of residual pairs also grows with laser intensity but $n_r \propto \nu^2$.  With optical lasers at the forefront of the current generation, these dynamical QED vacuum effects may generate $5-10$ observable two-photon annihilation events per laser pulse. 
\end{abstract}

%{\it Key words:\/} Pair production; Vacuum; Electron; Positron;
%Laser; Kinetic equation


\maketitle

%\textbf{1.} 
In the presence of a strong external electric field the vacuum of QED ``breaks down'' via the emission of electron-positron pairs \cite{SHE,JS}.  A theoretical understanding of this phenomenon is well established; e.g., Refs.\,\cite{Fradkin,Grib}, but hitherto an experimental verification is lacking.  A key obstacle is the very high value of the electric field required to achieve this phenomenon; namely, for electrons, $E_{cr} = m^2/e = 1.3 \times 10^{16}\,$V/cm.  (We use $\hbar = 1 = c$.)  According to Schwinger's formula \cite{JS}, the pair creation rate in a constant electric field is exponentially damped for $E\ll E_{cr}$.  However, a very different situation exists when the electric field is strongly time-dependent \cite{Grib,Smol,kme,Roba,Robb,Pop}.  In that case the Schwinger formula and its analogue for a monochromatic field become inapplicable in the weak-field regime \cite{Brezin}.  Despite the high value of $E_{cr}$, examples do exist of physical situations in which vacuum pair-production can occur, such as: relativistic heavy ion collisions \cite{Casher}; neutron stars \cite{Ruffini}; and focused laser pulses \cite{focus}.  

A description of an electromagnetic field may be obtained using ${\cal F} = (\vec E^2 - \vec B^2)/2$, ${\cal G} = \vec E \cdot \vec B$.  No pairs can be produced when ${\cal F}=0={\cal G}$, which is the case for an electromagnetic plane wave.  This is also approximately true of the field produced by focused laser beams \cite{Troup}, in which case pair production is exponentially suppressed.  On the other hand, it should be possible to avoid the lightlike field configuration with a spatially uniform field created in an antinode of the standing wave produced by the superposition of two coherent, counter-propagating laser beams \cite{Marinov}.  Pair creation is a nonperturbative effect and no complete solution of the relevant dynamical equations is available for a realistic configuration of laser fields.  However, numerous studies exist for the idealised situation of spatially-uniform time-dependent fields \cite{Roba,Robb,Pop,Brezin,Casher,Ruffini,focus,Troup,Marinov,Bunkin,Avetissian,piazza,BulanovSS} with the conclusion that vacuum decay is not observable with the laser parameters currently available.

With recent developments in laser technology, in particular the method of chirped pulse amplification, having yielded a remarkable increase in light intensity at the laser focal spot \cite{BulanovSV}, and with the construction of X-ray free electron lasers (XFELs) now underway, the possibility of an experimental verification of spontaneous pair creation from the vacuum is again attracting attention \cite{Ring,mckellar}.  

Vacuum decay is a far-from-equilibrium, time-dependent process and hence kinetic theory provides an appropriate descriptive framework.  We employ the approach of Ref.\,\cite{Smol}, which allows one to consider pair production as a dynamical process while accounting properly for the initial conditions.  This method is essentially nonperturbative and possesses novel features.  For example, it incorporates the essentially non-Markovian character of pair production in quantum field theory and its dependence on particle statistics \cite{our,quark}, and provides for a description of the complete momentum-dependence of the single-particle distribution function.  A characteristic feature of the kinetic approach is an ability to describe quasiparticle excitations during all stages in the evolution of an external field.

This quantum kinetic framework was used in Refs.\,\cite{Roba,Robb} to study an electric field with near critical magnitude and X-ray frequency.  It was shown that a field magnitude of approximately $0.25 E_{cr}$ could initiate particle accumulation and the consequent formation of a plasma of spontaneously produced pairs.  The quantum Vlasov equation of Ref.\,\cite{Smol} has also been employed in studies of the pre-equilibrium phase in the evolution of a quark-gluon plasma, whose creation on earth via ultrarelativistic heavy ion collisions is an aim at RHIC and LHC \cite{bastirev}.  

Herein, on the other hand, we consider the possibility of pair production with field parameters that are achievable today at laser facilities which are already in operation \cite{Jena,SLAC}; namely, $\nu^2 \ll E\ll E_{cr}$, where $\nu$ is the laser field frequency.  As gauges of creation efficiency we employ the mean density per period, $\langle n\rangle$, and the residual density taken over an integer number of field periods, $n_r$ \cite{Pop}.  We argue that, in comparison with XFELs, modern optical lasers can generate more vacuum polarisation $e^+ e^-$ pairs owing to the larger spot volume $\sim \lambda^3$, where $\lambda$ is the wavelength of the laser light, and hence may provide access now to observable signals of vacuum decay, such as coincident photon pairs from $e^+ e^-$ annihilation. 

%\textbf{2.} 
The key quantity in our approach is the single-particle momentum distribution function $f(\mathbf{p},t)$.  The kinetic equation satisfied by $f(\mathbf{p},t)$ may be derived from the Dirac equation in an external time-dependent electric field via the canonical Bogoliubov transformation method \cite{Grib}, or with the help of an oscillator representation \cite{OR}.  These procedures are only valid for simple field configurations; e.g., a spatially uniform, time-dependent electric field $\mathbf{E}(t)=(0,0,E(t))$, which is the idealisation we shall consider.  The field is assumed to vanish at an initial time $t = t_0$, whereat real particles are absent.  This is the ground state.  Ignoring collisions, which experience informs us is valid for the relatively weak field strengths considered herein \cite{Roba,Robb}, then $f(\mathbf{p},t)$ satisfies \cite{Smol}
\begin{eqnarray}
\nonumber
\frac{\partial f(\mathbf{p},t)}{\partial t} &+& e
\mathbf{E}(t)\frac{\partial f(\mathbf{p},t)}{\partial
\mathbf{p}} = \frac12 \Delta(\mathbf{p},t,t)\int\limits_{t_0}^t \! dt_1 \,
\Delta(\mathbf{p},t_1,t)\\
&\times&  \left[ 1- 2 f(\mathbf{p},t_1)\right] \cos [x(t,t_1)],
\label{ke}
\end{eqnarray}
where the three-vector momentum $\mathbf{p} = (\mbox{\boldmath $p$}_\perp,p_\|)$ and
\begin{eqnarray}
\mathbf{p}(t_1,t_2) &= &\mathbf{p} - e\int\limits_{t_1}^{t_2} \mathbf{E} (t')dt', \\
\Delta(\mathbf{p},t_1,t_2)& = & \frac{e E(t_1) \epsilon_\perp}
{\varepsilon^2(\mathbf{p},t_1,t_2)}\,, \\
%
%\varepsilon(\mathbf{p},t_1,t_2)&
%= &\sqrt{\epsilon_\perp^2 +
%p^2_{\mbox{\tiny$\parallel$}}(t_1,t_2)}\,,\\
%
x(t,t_1) & = & 2 \int\limits_{t_1}^{t}dt_2\,\varepsilon(\mathbf{p},t_2,t),
\end{eqnarray}
$\varepsilon^2(\mathbf{p},t_1,t_2) 
= \epsilon_\perp^2 + p^2_{\mbox{\tiny$\parallel$}}(t_1,t_2)$,
with $\epsilon_\perp^2 = m^2+p_\perp^2$.
%
%with $m$ the electron mass.

The total field $E(t)$ is defined as the sum of the external (laser) field $E_{ex}$ and the self-consistent internal field $E_{in}$, which is determined by Maxwell's equation
\begin{eqnarray}
\nonumber 
\lefteqn{\dot E_{in}(t) =  - e \int\frac{d^3 p}{(2\pi)^3} 
\frac{1}{\varepsilon_0}\bigg[ 2 p_\| f(\mathbf{p},t)  }\\
& & + \epsilon_\perp \int\limits_{t_0}^t \! dt_1 \Delta(\mathbf{p},t_1,t) \left[ 1- 2 f(\mathbf{p},t_1)\right] 
\cos [x(t,t_1)] \rule{0em}{4ex}\bigg], \label{max}
\end{eqnarray}
where $\varepsilon_0 = \varepsilon(\mathbf{p},t,t)$.  The current density on the
r.h.s.\ of Eq.~(\ref{max}) is the sum of a conduction current, proportional to $f(\mathbf{p},t)$ and tied to the particles' motion, and a polarisation current, linked to the pair production rate.

\begin{figure}[t]
\includegraphics[width=0.45\textwidth]{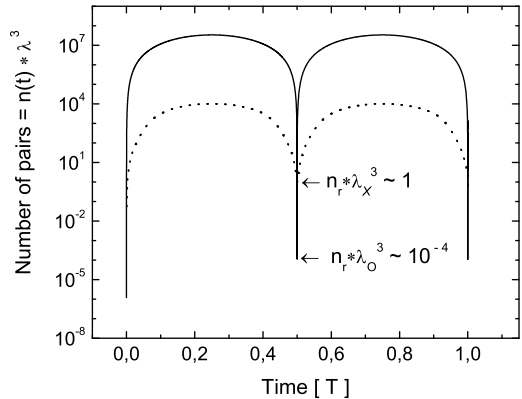}
\caption{$\lambda^3 n(t)$ as a function of time, measured in units of the laser period, $T$; i.e., the number of pairs produced within a volume $\lambda^3$ by the field in Eq.\,(\protect\ref{harm}). \emph{Solid line} -- optical laser (weak field case) \cite{Jena}: $E_m^{O} = 3\times 10^{-5} E_{cr}$ and $\lambda_{O} = 795\,$nm.  \emph{Dotted line} -- XFEL (strong field) \cite{Ring,Roba,Robb}: $E_m^X = 0.24\, E_{cr}$ and $\lambda_X = 0.15\,$nm.  The value of the residual pair density, $n_r$, is marked in both cases.  NB.\ $(\lambda_O/\lambda_X)^3= 1.5 \times 10^{11}$. \label{log}}
\end{figure}

Equation (\ref{ke}) is an integro-differential equation.  Its solution is complicated by the presence of three time-scales, which can be vastly different: $\tau_{\rm qu}=1/m$, the quantum time-scale that expresses intrinsically quantum field theoretic effects; $\tau_{\rm tu}=m/(e E)$, the time-scale characterising the separation between pair production events; and $\tau_{\ell}=1/\nu$, the laser period.  For the fields of interest herein $\tau_{\rm tu} \gg \tau_{\rm qu}$ and $\tau_{\ell} \gg \tau_{\rm qu}$.  However, despite this appearance of two small parameters, one cannot develop a perturbative solution because of the temporal nonlocality expressed  in the source via the coherent phase oscillation term: $\cos[x(t,t^\prime)]$.  Simplification is nevertheless possible because, with optical-laser-like parameters, $E \ll E_{cr}$ and consequently \cite{Roba,Robb} the quasiparticle number density is small; viz.,  $[1- 2 f(\mathbf{p},t)] \approx 1$, and the internal field $E_{in}$ is negligible.  Under these conditions, the solution is %of the kinetic equation is
\begin{eqnarray}
\nonumber\lefteqn{f(\mathbf{p},t)=
\,\frac12\, 
\int\limits_{t_0}^t dt_1 \,\Delta(\mathbf{p},t_1,t)}\\
&\times & 
\int\limits_{t_0}^{t_1} dt_2 \,\Delta(\mathbf{p},t_2,t)\,
\cos[x(t,t_2)]\,, \label{fact}
\end{eqnarray}
%\begin{multline}\label{fact}
%f(\mathbf{p},t) = \left| \,\frac12\, \int\limits_{t_0}^t dt_1 \,
%\Delta(\mathbf{p},t_1,t)\right.\\ \left. \times
%\exp{\left(2i\int\limits^{t_1}_{t_0} dt_2 \varepsilon
%(\mathbf{p},t_2,t)\right)}\right|^{\,2} \,.
%\end{multline}
%
from which the result is obtained directly via numerical integration subject to the initial condition $f(\mathbf{p},t_0)=0$.  The number density is 
\begin{equation}\label{dens}
n(t)=2 \int\frac{d^3 p}{(2\pi)^3} \, f(\mathbf{p},t)\ .
\end{equation}

We consider herein a simple model for the field formed in the superposition of two coherent, counter-propagating laser beams; i.e., an harmonic field, with magnitude $E_m$ and angular frequency $\omega = 2\pi \nu$, that persists for $z$ periods of length $T=1/\nu$:  
\begin{equation}\label{harm}
E(t) = E_m \sin{\omega t},\qquad 0 \le t \le z T.
\end{equation}

In Fig.\,\ref{log} we plot the time dependence of the quasiparticle pair density generated by fields of the type in Eq.\,(\ref{harm}).  Two field strengths are considered: one that represents the parameters of a working Ti:sapphire laser \cite{Jena}, with $E_m^O \approx 3\times 10^{-5} E_{cr}$ and $\lambda_O = 1/\nu = 795\,$nm; and another which mimics the planned XFEL at DESY \cite{Ring}, with $E_m^X = 0.24\, E_{cr}$ and $\lambda_X = 0.15\,$nm.  It is apparent that the density of $e^+ e^-$ quasiparticle pairs oscillates in tune with the field frequency \cite{Roba,Robb}.

\begin{figure}[t]
\includegraphics[width=0.45\textwidth]{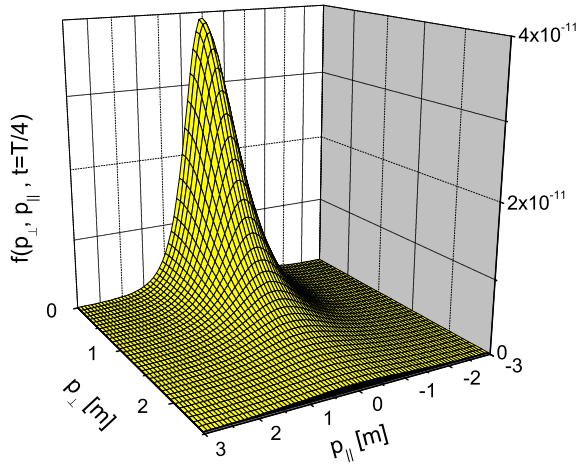}
\caption{\label{sp} Single particle momentum distribution function at an antinode of the electric field in Eq.\,(\protect\ref{harm}) for optical laser parameters; viz., $E_m= 3\times 10^{-5}\, E_{cr}$.} 
\end{figure}

We now introduce the residual and mean pair densities:
\begin{equation}
\label{timeaverage}
n_r := n( z T)\,, \; \langle n \rangle := \frac{1}{z T}\int_{t_0}^{t_0 + z T} \!\!\!dt\, n(t)\,.
\end{equation}
For fields of the type in Eq.\,(\ref{harm}), one finds
\begin{equation}
\label{nr}
 \lambda^3 \langle n \rangle \sim \left[\frac{e E_m}{m^2}\right]^2 \left[\frac{m \lambda}{2\pi}\right]^3 \!,\;\; \frac{n_r}{\langle n \rangle} \sim \frac{\omega^2}{m^2}\,,
\end{equation}
and for optical lasers, as one would anticipate, this ratio is very small.  For example (see Fig.\,\ref{log}), with the model optical laser parameters $n_r \sim 10^{-4} \lambda_O^{-3}$, $\langle n \rangle \sim 10^7 \lambda_O^{-3}$ and $n_r/\langle n \rangle \simeq  10^{-11}$; whereas for the XFEL parameters $n_r \sim \lambda_X^{-3}$, $\langle n \rangle \sim 10^4 \lambda_X^{-3}$ and $n_r/\langle n\rangle \sim 10^{-4}$.  

On the other hand, these results reveal that despite the fact that the residual density under XFEL conditions exceeds that of optical lasers by many orders of magnitude, the number of $e^+ e^-$ quasiparticle pairs within the spot volume is far greater for optical lasers.  Indeed, on average, optical lasers produce roughly $10^7$ virtual pairs in their spot volume during each laser period.  This corresponds to a vacuum polarisation pair density of $\sim 10^{20}\,$cm$^{-3}$; i.e., a dense plasma of $e^+ e^-$ quasiparticle pairs that vanishes almost completely at the field's nodal points.  NB.\ This outcome is readily understood: the spot volume for optical lasers is much larger than that for a typical XFEL.

One may compare our result with Ref.\,\cite{Pop}, which employs an imaginary time method that yields $n_r \sim z$ but no information about $\langle n \rangle$.  In Eq.\,(\ref{nr}) we report that the mean density of $e^+ e^-$ quasiparticle pairs is independent of $\nu$, while $n_r \sim \nu^2$.  Both densities are proportional to the laser's intensity and this leads to the accumulation effect for $n_r$ in near critical fields \cite{Robb}.  For subcritical fields, the number of $e^+ e^-$ pairs remaining after an integer number of periods is negligible in comparison with the mean density. %$^1$ \footnotetext[1](
(NB.\ Eq.\,(\ref{nr}) is not applicable for pulse-shaped fields, which may be a more realistic model for crossed lasers.  For this geometry $n_r$ depends strongly on the parameters that determine the pulse shape but this is not material to our subsequent discussion, which is based on results determined numerically.)
%}

\begin{figure}[t]
\includegraphics[width=0.45\textwidth]{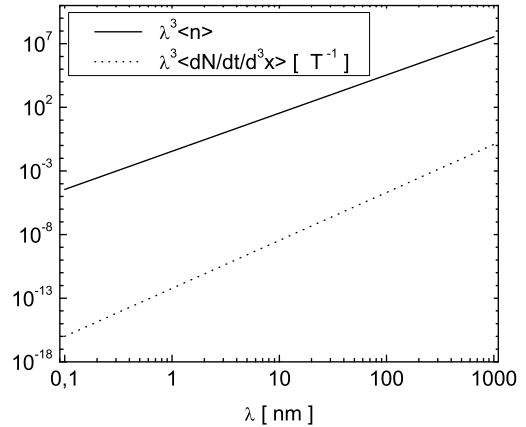}
\caption{\emph{Solid line} -- $\lambda$-dependence of the mean quasiparticle number, $\lambda^3 \langle n \rangle $; and \emph{dotted line} -- $\gamma\gamma$-production-rate/laser-period from the spot volume, $\lambda^3 \langle d N_{e^+ e^-}/dt/ d^3x\rangle $ (with the time-averaging procedure defined in Eq.\,(\protect\ref{timeaverage}).  Both curves were calculated with a fixed field strength $E_m= 3\times 10^{-5}\, E_{cr}$ in Eq.\,(\protect\ref{harm}). \label{freq}}
\end{figure}

In Fig.\,\ref{sp} we plot the single-particle momentum distribution at an antinode of our model for an optical laser field.  Consistent with Ref.\,\cite{Roba}, even for this weak field the distribution function has a longitudinal and transverse momentum space width $\sim m$.  This is in contrast to a common assumption that the longitudinal momentum of the produced pairs vanishes \cite{Casher}.

We have shown that an optical laser can induce a significant polarisation of the QED vacuum.  To determine whether this has observable consequences we estimate the intensity of $e^+ e^- \to \gamma \gamma$ annihilation from the polarisation volume.  The $\gamma \gamma$ signal, with mean total energy $\approx 1\,$MeV (cf.\ the laser photon energy of $\sim 1\,$eV), should be seen outside the laser spot volume.  The $\gamma\gamma$ rate is
\begin{eqnarray}
\nonumber 
\lefteqn{\frac{d N_{e^+ e^-}}{dt\, d^3x}= \int\frac{d^3 p_1}{(2\pi)^3}\frac{d^3 p_2}{(2\pi)^3} f_1(\mathbf{p}_1,t)\, f_2(\mathbf{p}_2,t)} \\
 & \times&  \sigma(\mathbf{p}_1,\mathbf{p}_2) \,   
\sqrt{(\mathbf{v}_1-\mathbf{v}_2)^2 - (\mathbf{v}_1 \times \mathbf{v}_2)^2}\,,
\label{num}
\end{eqnarray}
where $\mathbf{v}$ is a particle's velocity and $\sigma$ is the cross-section for two-photon annihilation
\begin{eqnarray}
\nonumber
\lefteqn{\sigma(\mathbf{p}_1,\mathbf{p}_2) =  \frac{\pi e^4}{2m^2 \hat t^2
(\hat t-1)} \biggl[ \bigl(\hat t^2 + \hat t - 1/2 \bigr)}\\
&& \times \ln{\left\{ \frac{\sqrt{\hat t} +
\sqrt{\hat t-1}}{\sqrt{\hat t} - \sqrt{\hat t-1}}\right\} - (\hat t+1)
\sqrt{\hat t (\hat t-1)} } \biggr],
\label{sigma}
\end{eqnarray}
with the t-channel kinematic invariant 
\begin{equation}\label{tau}
\hat t = \frac{(p_1 + p_2)^2}{4m^2} = \frac{1}{4m^2}\bigl[
(\varepsilon_1 + \varepsilon_2)^2 - (\mathbf{p}_1 +
\mathbf{p}_2)^2 \bigr].
\end{equation}

For this estimate, we consider a laser-induced field specified by the following parameters \cite{Jena}: pulse intensity $I=10^{20}$ W/cm$^2$; pulse duration $\tau_L = 85\,$fs and $\lambda=795\,$nm; and spot diameter $2.5\mu$m, and find there are $5 -10$ annihilation events per laser pulse.  We depict the wavelength dependence of the mean particle number and spot-volume production-rate in Fig.\,\ref{freq}.  Plainly, as noted in Ref.\,\cite{Robb}, if all other factors can be maintained, there is merit in increasing $\lambda$.  While more dramatic signals must likely await XFEL capacities \cite{Roba,Robb}, this study suggests the intriguing possibility that contemporary laser facilities may be sufficient for the first observation of an intrinsically nonperturbative effect in QED.

%\textbf{4.} 
We explored the possibility of $e^+ e^-$ pair production using the present generation of optical lasers %\cite{Jena,SLAC} 
as a parameter-free application of non-equilibrium quantum mean field theory.  
With an idealised model for a crossed-laser electric field as input to a quantum Vlasov equation, we found a significant polarisation of the QED vacuum.  It is characterised by a dense plasma of $e^+ e^-$ quasiparticle pairs,  which disappears almost completely once the laser field vanishes, leaving a very small residual pair density.  The mean density is independent of the laser frequency, $\nu$, while the density of residual pairs increases with $\nu^2$.  
%
%We argued that 
These dynamical QED vacuum effects may be signalled by the appearance of coincident photon pairs, from $e^+ e^-$ annihilation, with a mean energy of $\sim 1\,$MeV and an intensity of $5-10$ events per laser pulse.  
This represents a nonlinear transformation of soft laser photons to $\gamma$-quanta with a frequency ratio of $\gtrsim 10^6$.

%\begin{acknowledgments}
We thank D.~Habs, A.~H\"oll, P.~Jaikumar, \mbox{P.-V.~Nick}\-les, G.~R\"opke, M.~Romanovsky, R.~Sauerbrey and S.\,V.~Wright for useful discussions.  
This work was supported by: 
Department of Energy, Office of Nuclear Physics, contract no.\ W-31-109-ENG-38; 
%
%\textit{Helmholtz-Gemeinschaft} Virtual Theory Institute VH-VI-041; 
%
and the \textit{A.\,v.\ Humboldt-Stiftung} via a \textit{F.\,W.\ Bessel Forschungspreis}.
%\end{acknowledgments}

\end{document}